\documentclass[a4paper,11pt]{article}
\pdfoutput=1 
\usepackage{jcappub}
\usepackage{latexsym} 
\usepackage{graphicx}
\usepackage{pxfonts}
\usepackage{xcolor}

\title{\boldmath Images of black holes viewed by distant observer}
\author[a,1]{V. I. Dokuchaev\note{Corresponding author.}}
\affiliation[a]{Institute for Nuclear Research, Russian Academy of Sciences, \\
	60th October Anniversary Prospect 7a, 117312 Moscow, Russia}
\emailAdd{dokuchaev@inr.ac.ru}
\abstract{We describe the possible forms of black hole images, viewed by a distant observer. These images are numerically calculated basing on general relativity and equations of motion in the Kerr-Newman metric. Black hole image is a gravitationally lensed image of the black hole event horizon. It may be viewed as a black spot on the celestial sphere, projected inside the position of classical black hole shadow. In the nearest future it would be possible to verify modified gravity theories by observations of astrophysical black holes with Space Observatory Millimetron.} 
\begin{document}
	\maketitle
	\flushbottom

\section{Introduction}
\label{intro}

How black hole looks like? It is a standard question of both scientific experts and men in the streets. In this paper the black hole images are calculated basing on general relativity and equations of motion in the classical  Kerr--Newman metric~\cite{Kerr,Newman1,Newman2, Carter68,  mtw,Chandra}, describing the rotating and electrically charged black hole (see appendix~\ref{KN}, \ref{Eqs} and \ref{Int} for details).  These images are the gravitationally lensed images of black hole event horizon.   

Numerical supercomputer simulations of general relativistic hydro-magnetic accretion onto black (see,  e.\, g., \cite{Tchekhovskoy11, Tchekhovskoy12,Tchekhovskoy12b, Tchekhovskoy15, Tchekhovskoy17,Tchekhovskoy17b, Ryan18}) affirm the Blandford-Zna\-jek mechanism \cite{BlandfordZ} of energy extraction from fast  rotating Kerr black holes. The crucial feature of this mechanism is an electric current flowing through the black hole immersed into external poloidal magnetic field. This electric current heats the accreting plasma up to the nearest outskirts of the black hole event horizon. Very high luminosity of this hot accreting plasma will spoil some parts of the  black spots at the astrophysical black hole images. 

Images of astrophysical black holes may be viewed as black spots on the celestial sphere, projected inside the possible positions of classical black hole shadows.  See at figure~\ref{fig1} and figure~\ref{fig2} some examples of the classical black hole shadows.

It must be stressed that the forms of discussed dark spots are independent on the distribution and emission of the accreting plasma. Instead of, the corresponding forms of dark spots are completely defined by the properties of black hole gravitational field and black hole parameters like black hole mass $M$ and spin $a$. (Throughout this paper we use the standard dimensionless units with $GM/c^2=1$, where $G$ --- Newtonian  gravitational constant, $c$ --- velocity of light). 

See at figure~\ref{fig3} the example of reconstruction of the spherically symmetric Shwarzschild black hole event horizon silhouette using $3D$ trajectories of photons (multicolored curves), which start very near the black hole event horizon and are registered by a distant observer (by a distant telescope). Correspondingly, at figure~\ref{fig4} is shown the example of reconstruction of the extremely fast rotating Kerr black hole with spin $a=1$.

The trajectories of photons at all Figures of this paper are calculated numerically by using test particle equations of motion in the Kerr metric (see appendix \ref{Eqs} and \ref{Int}). The event horizon silhouette (dark spot) always projects at the celestial sphere within the classical black hole shadow.

\section{Classical black hole shadow}

\begin{figure}
	\begin{center}
		\mbox{\includegraphics[width=0.51\textwidth]{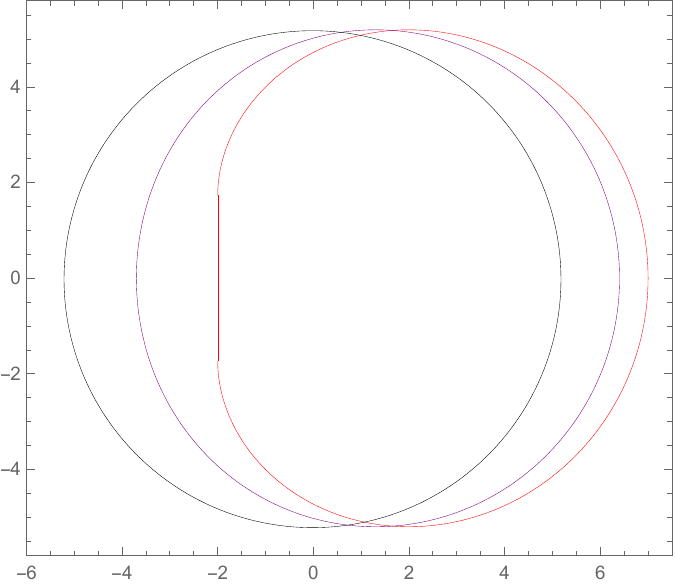}}
		\hfill
		\makebox{\includegraphics[width=0.44\textwidth]{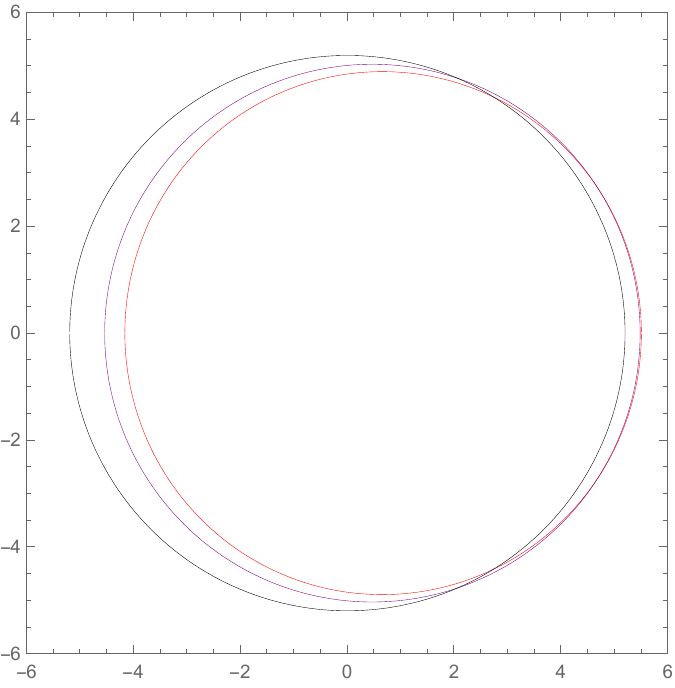}}	
		\caption{Some examples of classical black hole shadows are shown for the cases of supermassive black holes SgrA* and M87*. Left panel: the shadows of SgrA* (with a possible inclinations of rotation axes with respect to the polar angle $\theta_0$) in the spherically symmetric Schwarzschild case $(a=0)$ is black circle with a radius $r_{\rm sh}=3\sqrt{3}\simeq5.196$.  The closed {\color{red} red} curve is the shadow of extremely fast rotating black hole $(a=1)$ and closed {\color{purple} purple} curve $(a=0.65)$, respectively. Note, that the vertical sizes of shadows in the case of SgrA* are independent on the values of spin $a$. Right pane: The corresponding forms of shadows in the case of M87* (with a possible inclinations of rotation axes with respect to the polar angle $\theta_0=163^\circ$) for spin values $a=1$ ({\color{red} red} closed curve), $a=0.75$ ({\color{purple} purple} closed curve), and $a=0$ (black circle) of radius $r_{\rm sh}=3\sqrt{3}$.}
		\label{fig1}
	\end{center}
\end{figure}

Some examples of the classical black hole shadows are shown at figure~\ref{fig1} for the cases of supermassive black holes M87* at the center of galaxy M87 and SgrA* at the center of our native Milky Way galaxy. Left panel: the shadows of SgrA* (with a possible inclinations of rotation axes with respect to the polar angle $\theta_0$) in the spherically symmetric Schwarzschild case $(a=0)$ is black circle with a radius $r_{\rm sh}=3\sqrt{3}\simeq5.196$.  The closed {\color{red} red} curve is the shadow of extremely fast rotating black hole $(a=1)$ and closed {\color{purple} purple} curve $(a=0.65)$, respectively. Note, that the vertical sizes of shadows in the case of SgrA* are independent on the values of spin $a$. Right pane: The corresponding forms of shadows in the case of M87* (with a possible inclinations of rotation axes with respect to the polar angle $\theta_0=163^\circ$) for spin values $a=1$ ({\color{red} red} closed curve), $a=0.75$ ({\color{purple} purple} closed curve), and $a=0$ (black circle) of radius $r_{\rm sh}=3\sqrt{3}$.

\begin{figure}
	\begin{center}
		\includegraphics[width=0.85\textwidth]{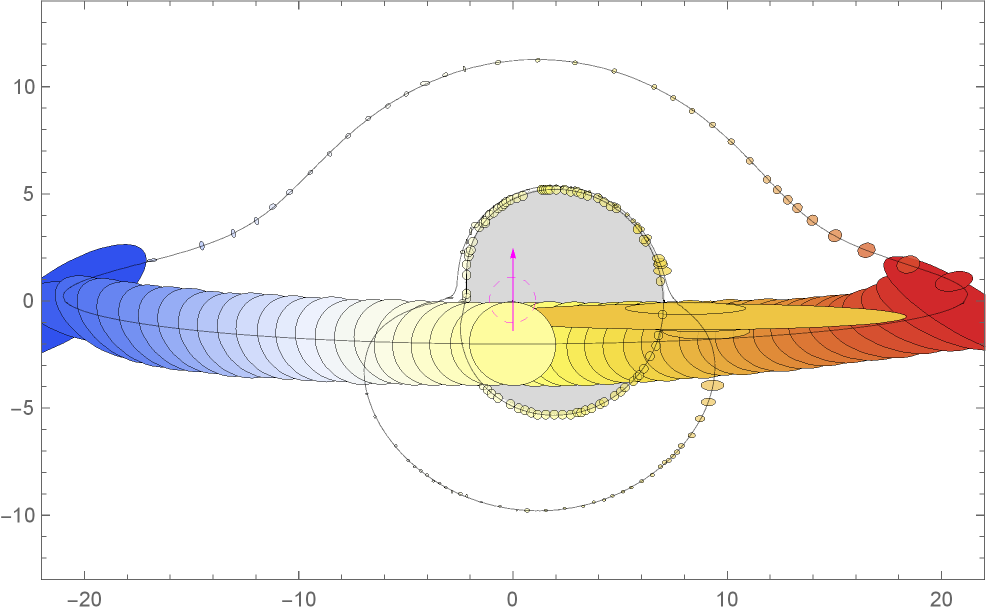}
		\caption{The numerical simulation of compact spherical probe (neutron star or spaceship) orbiting around a fast-rotating black hole ($a=0.9982$) at a circular orbit with a dimensionless radius $r=20$.  It is shown one orbital period in discrete time intervals. A distant observer is placed a little bit above the black hole equatorial plane. It is shown the direct image and the first and second light echoes. The {\color{gray} gray} region is classical black hole shadow. The images second light echo is concentrated at the outskirts of shadow. It is also considered the gravitational lensing of spherical probe in the black hole gravitational field (in the ellipsoidal approximation), viewed by distant observer as deformation of the probe.  For details see \cite{doknaz18b}.} 
		\label{fig2}
	\end{center}
\end{figure}

\begin{figure}
	\begin{center}
		\includegraphics[width=0.95\textwidth]{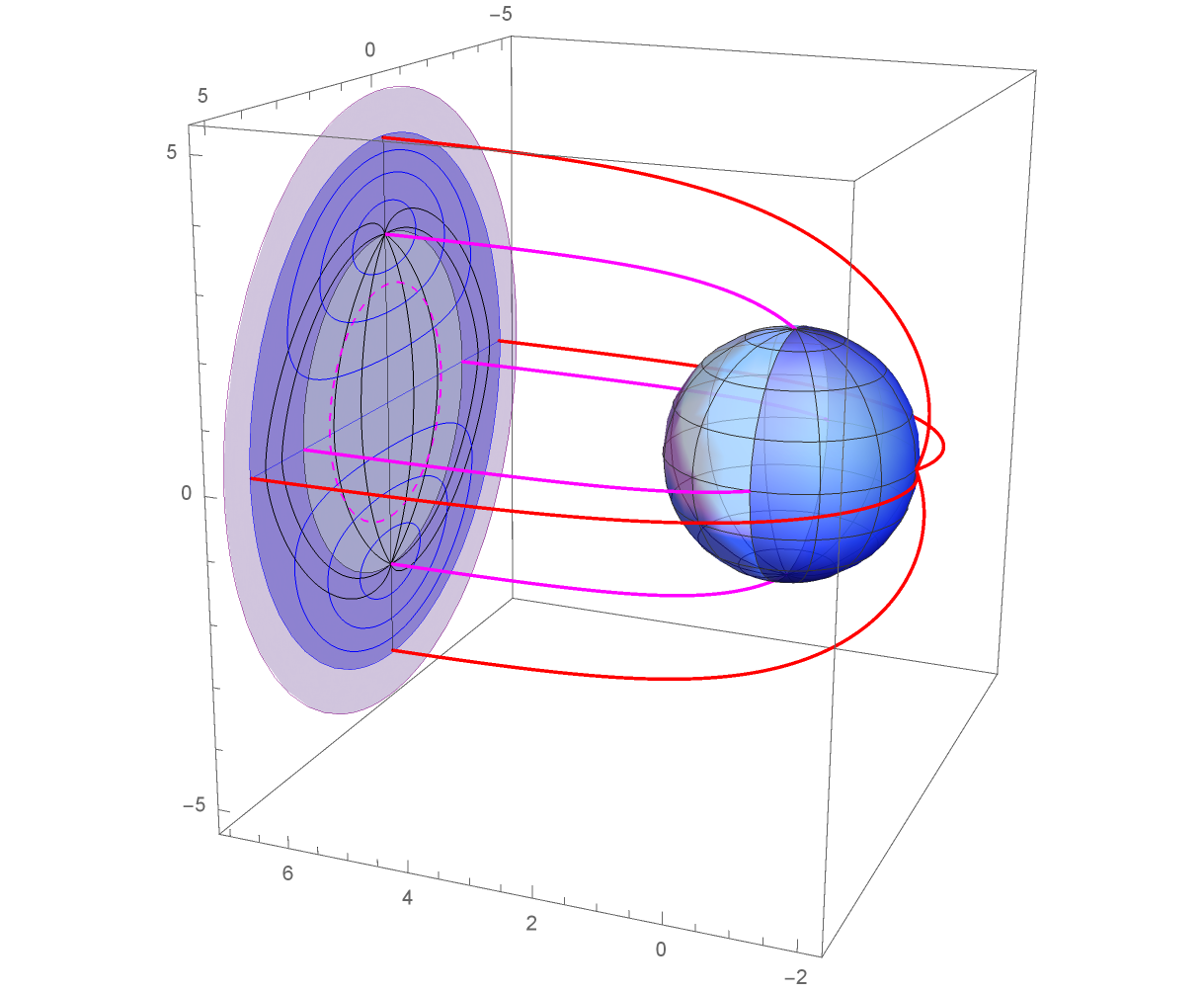} 
		\caption{Reconstruction of the Shwarzschild black hole event horizon silhouette using $3D$ trajectories of photons (multicolored curves), which start very near the black hole event horizon and are registered by a distant observer (by a distant telescope). The event horizon silhouette always projects at the celestial sphere within the classical black hole shadow with a radius $3\sqrt{3}$. Meanwhile, the corresponding radius of event horizon silhouette is $r_{\rm h}\simeq4.457$ \cite{doknaz19,doknaz20}.}
		\label{fig3}
	\end{center}
\end{figure}

\begin{figure}%
	\begin{center}
		\includegraphics[width=0.8\textwidth]{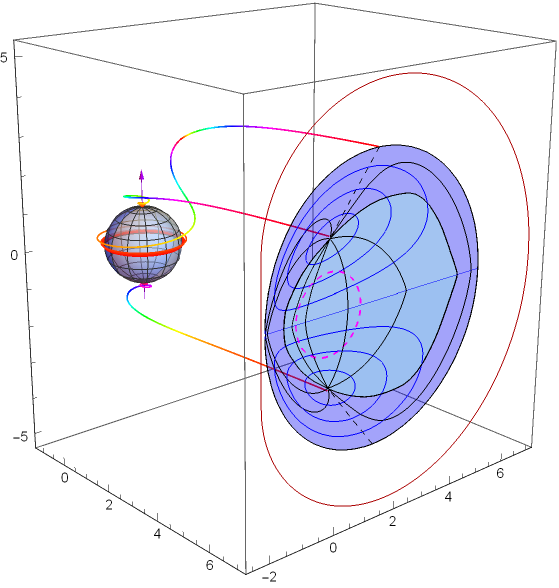}
		\caption{Reconstruction of extremely fast Kerr black hole ($a=1$) event horizon silhouette for the rotation axes orientation of the supermassive black hole SgrA*.  It is used $3D$ trajectories of photons (multicolored curves), which start very near the black hole event horizon and are registered by a distant observer (by a distant telescope). The nearest hemisphere of the event horizon is projected into (light {\color{blue} blue}) disk with radius $r_{\rm EW}\simeq2.848$. The farthest hemisphere is projected into the hollow (dark {\color{blue} blue}) disk with radius $r_{\rm eh}\simeq4.457$. It is a radius of the gravitationally lensed event horizon image. The {\color{blue} blue} and black curves are the corresponding parallels and meridians at the event horizon globe.}
		\label{fig4}    
	\end{center}  
\end{figure}

See at figure~\ref{fig2} the numerical simulation of compact spherical probe (neutron star or spaceship) orbiting around a fast-rotating black hole ($a=0.9982$) at a circular orbit with a dimensionless radius $r=20$. It is shown one orbital period in discrete time intervals. A distant observer is placed a little bit above the black hole equatorial plane. It is shown the direct image and the first and the second light echoes. The {\color{gray} gray} region is the classical black hole shadow. The images second light echo is concentrated at the outskirts of the shadow. It is also considered the gravitational lensing of spherical probe in the black hole gravitational field (in the ellipsoidal approximation), viewed by distant observer as deformation of the probe.  For details of numerical calculations see \cite{doknaz18b}.

\begin{figure}
	\begin{center}
		\includegraphics[width=0.65\textwidth]{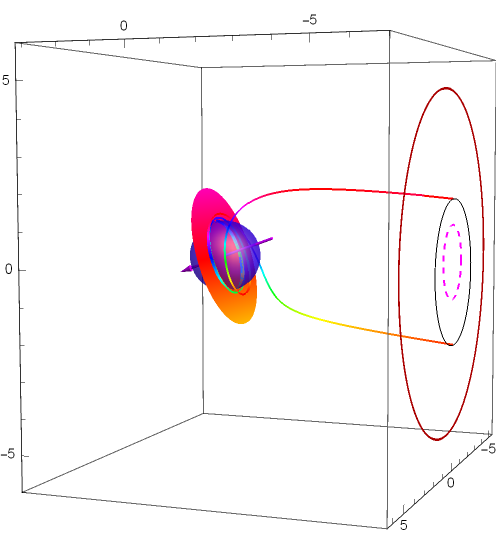} 
		\caption{$3D$ picture of the supermassive black hole  M87* with a supposed spin parameter $a=1$ surrounded by thin accretion disk, which is supposed to be nontransparent. The closed  {\color{gray}gray} curve is an outer boundary of the dark spot. Two numerically calculated (multicolored) photon trajectories are started from the inner boundary of the accretion disk (in the vicinity of black hole event horizon equator) and finished far from black hole at the position of a distant observer. The largest closed  {\color{purple}purple}  curve at this  $3D$ picture is an outer boundary of the classical black hole shadow.}
		\label{fig5}
	\end{center}
\end{figure}

\begin{figure}
	\begin{center}
		\includegraphics[angle=0,width=0.75\textwidth]{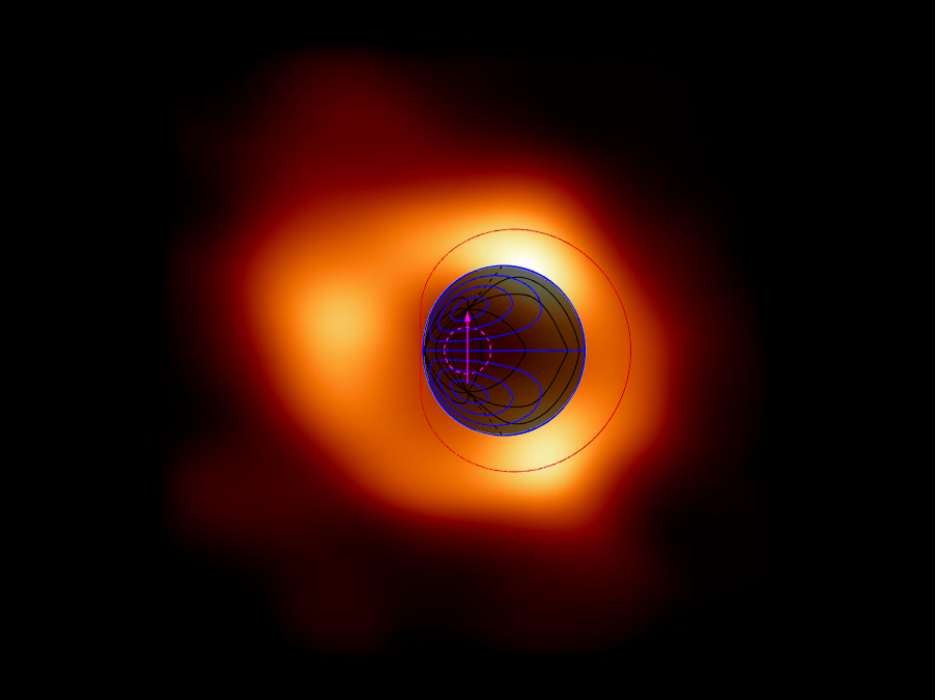}
		\caption{A composition of the Event Horizon Telescope image of supermassive black hole SgrA* \cite{EHT1,EHT2,EHT3,EHT4,EHT5,EHT6, EHT1S} with the  numerically modeled dark spot, corresponding to the gravitationally lensed image of the event horizon globe with spin $a=1$. The closed purple curve is the outline of classical black hole shadow. The magenta arrow is the direction of black hole rotation axis. The magenta dashed circle is a position of the black hole event horizon with radius $r_{\rm h}=1$ in the Euclidean space without gravity.} 
		\label{fig6}
	\end{center}
\end{figure}

\begin{figure}
	\begin{center}
		\mbox{\includegraphics[width=0.48\textwidth]{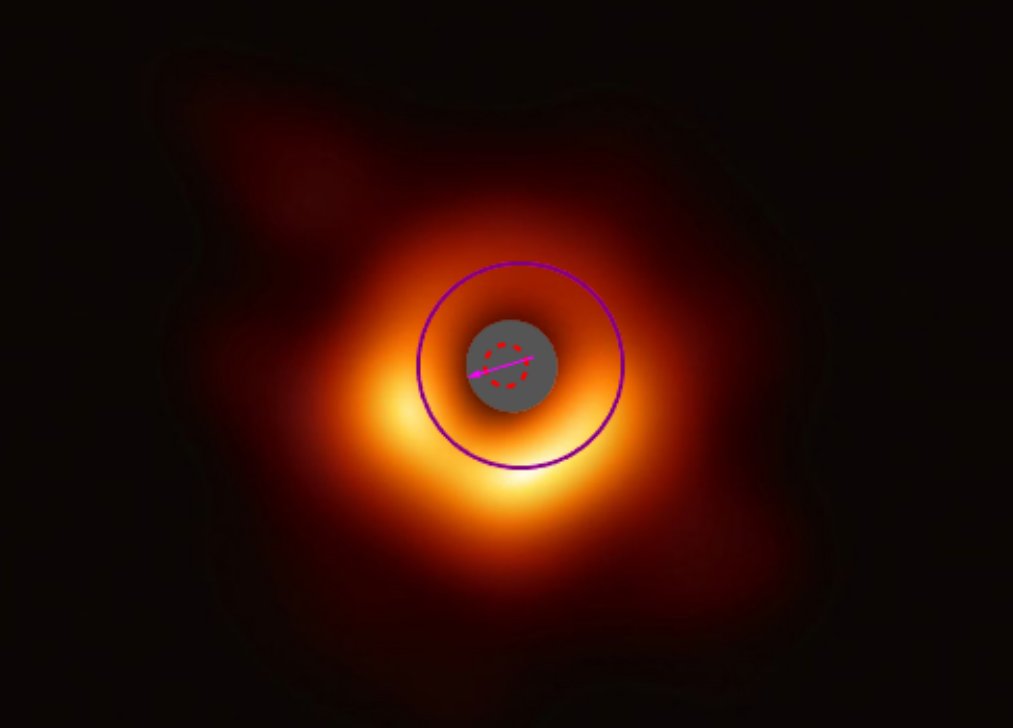}}
		\hfill
		\makebox{\includegraphics[width=0.48\textwidth]{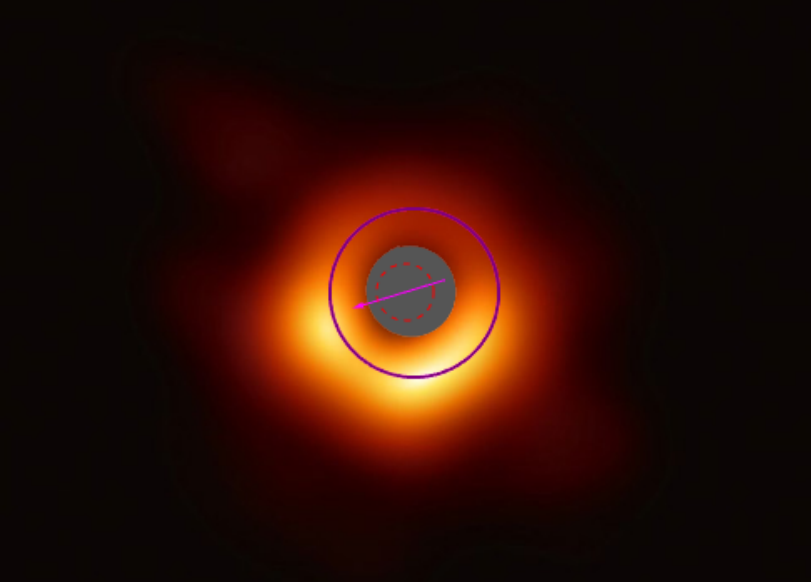}}
		\caption{Superposition of the modeled dark spot with the Event Hotizon Telescope image of supermassive black hole M87*: Left panel, \ $a=1$; \ right panel, \ $a=0.75$. A $17^\circ$  inclination angle of the black hole rotation axis at the celestial sphere is supposed. Note that the size of dark spot in the case of rotation axis orientation of this black hole is weakly depends on the value of spin parameter $a$.}
		\label{fig7}
	\end{center}
\end{figure}

\begin{figure}
	\begin{center}
		\includegraphics[width=0.85\textwidth]{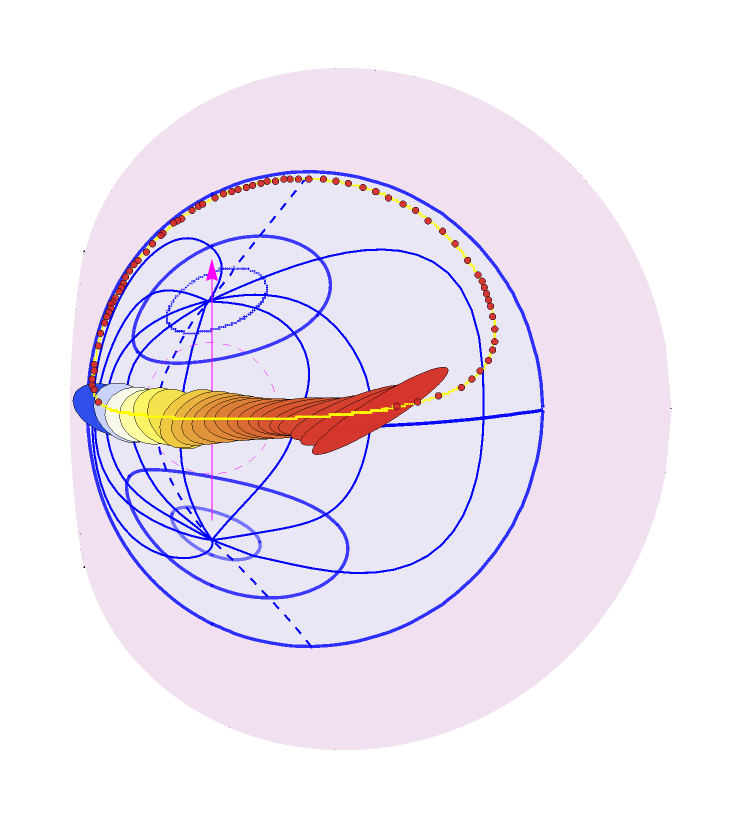}
		\caption{Gravitationally lensed images in discrete time intervals of small probe (neutron star or cosmic ship) with a zero angular momentum ($\lambda=0$) and zero Carter constant ($q=0$), which is plunging into a fast-rotating black hole. A distant observer is placed a little bit above the black hole equatorial plane. The small probe is winding around the black hole equator of the event horizon globe. It is shown one circle of this winding. The escaping signals from this prob are exponentially fading in time. For details of numerical calculations and for animation see \cite{dokuch14,dokuch19,doknaz19b,doknaz21, doknazsm19,Dokuch22}.}    
		\label{fig8}
	\end{center}
\end{figure}

The apparent  shape of the black hole shadow, as seen by a distant observer in the equatorial plane, is determined parametrically, $(\lambda,q)=(\lambda(r),q(r))$, from simultaneous solution of equations $V_r(r)=0$ and $[rV_r(r)]'=0$ (see  e.\, g., \cite{Bardeen73,Chandra,BisnovatyiTsupko18}):  
\begin{eqnarray} \label{shadow1}
	\lambda&=&\frac{-r^3+3r^2-a^2(r+1)}{a(r-1)}, \\
	q^2&=&\frac{r^3[4a^2-r(r-3)^2]}{a^2(r-1)^2}.
	\label{shadow2}
\end{eqnarray}

\section{Dark spots at black hole images}

The form of a dark spot at the astrophysical black hole image, which is viewed by a distant telescope (observer) at the black hole equatorial plane may be calculated by using Brandon Carter \cite{Carter68} integral equation of motion in the Kerr metric 
\begin{equation}
	\int_2^\infty\frac{dr}{\sqrt{V_r}}
	=2\int_{\theta_{\rm min}}^{\pi/2}\frac{d\theta}{\sqrt{V_\theta}}.
	\label{a0max}
\end{equation}
where $\theta_{\rm min}$ is a turning point of the photon trajectory for direct image in the polar direction (for details, see \cite{CunnBardeen72,CunnBardeen73}).

In the Schwarzschild case a turning point is at polar angle $\theta_{\rm min} = \arccos(q/\sqrt{q^2+\lambda^2})$, where $q$ and $\lambda$ are parameters of photon trajectories from equation (\ref{three}). Respectively, from the right-hand-side integral in (\ref{a0max}) is $\pi/\sqrt{q^2 +\lambda^2}$. The resulting numerical solution of integral equation (\ref{a0max}) provides the radius of event horizon image $r_{\rm eh}=\sqrt{q^2+\lambda^2}=4.457$. The nearest hemisphere of the event horizon is projected into (light {\color{blue} blue}) disk with radius $r_{\rm EW}\simeq2.848$. The farthest hemisphere is projected into the hollow (dark {\color{blue} blue}) disk with radius $r_{\rm eh}\simeq4.457$. It is a radius of the gravitationally lensed event horizon image.  At figure~\ref{fig3} and \ref{fig4} are shown the corresponding numerical solutions for the gravitationally lensed event horizon images in the Schwarzschild ($a=0$) and extremely fast Kerr rotation case ($a=1$) for the rotation axes orientation of the supermassive black hole SgrA*. The near hemisphere of the event horizon is projected by the lensing photons into the central (dark {\color{blue} blue}) region.  Respectively the far hemisphere is projected into the hollow (light {\color{blue} blue}) region. The {\color{blue} blue} and black curves are the corresponding parallels and meridians at the event horizon globe. 

The event horizon globe of the Kerr black hole ($e=0$,  $a\neq0$) according to general equation (\ref{omega+}) is rotating with an angular velocity as a solid body
\begin{equation}
	\Omega_{\rm h}=\frac{a}{2(1+\sqrt{1-a^2})}.
	\label{OmegaH}
\end{equation}

At figure~\ref{fig5} is shown a $3D$ picture of the supermassive black hole  M87* with a supposed spin parameter $a=1$ surrounded by thin accretion disk, which is supposed to be nontransparent. An inclination angle of M87* rotation axis with respect to a distant observer is supposed to be near $17^\circ$. The {\color{magenta}magenta} arrows indicate direction of the black hole rotation axis. The closed  {\color{gray}gray} curve is an outer boundary of the dark spot which may be viewed by a distant observer at the celestial sphere. Two numerically calculated (multicolored) photon trajectories are started from the inner boundary of the accretion disk (in the vicinity of black hole event horizon equator) and finished far from black hole at the position of a distant observer. The largest closed  {\color{purple}purple}  curve at this  $3D$ picture is an outer boundary of the classical black hole shadow. We remind that black spots on the images of astrophysical black holes are always projected inside the possible positions of the black hole shadows. The dashed {\color{red}red} circle is a projection on the celestial sphere of the imaginary sphere with unit radius in the absence of gravity (i/\,e. at the Euclidean space). The southern hemisphere of the gravitationally lensed event horizon globe may be viewed by a distant observer in the case of M87*. 

See at figure~\ref{fig8} the gravitationally lensed images in discrete time intervals of small probe (neutron star or cosmic ship) with a zero angular momentum ($\lambda=0$) and zero Carter constant ($q=0$), which is plunging into a fast-rotating black hole. A distant observer is placed a little bit above the black hole equatorial plane. The small probe is winding around the black hole equator of the event horizon globe. It is shown one circle of this winding. The escaping signals from this prob are exponentially fading in time. For numerical animation see \cite{doknaz18c}.

\section{Discussions and Conclusions}

In this paper the possible forms of black hole images, viewed by a distant observer, are calculated basing on general relativity and equations of motion in the Kerr-Newman metric. Black hole image is a gravitationally lensed image of the black hole event horizon. It may be viewed as a black spot on the celestial sphere, projected inside the position of classical black hole shadow. The event horizon silhouette (dark spot) always projects at the celestial sphere within the classical black hole shadow. 

Images of astrophysical black holes may be viewed as black spots on the celestial sphere, projected inside the possible positions of classical black hole shadows. Very high luminosity of hot accreting plasma will spoil some parts of the  black spots at the astrophysical black hole images. 

It must be stressed that the forms of discussed dark spots are independent on the distribution and emission of the accreting plasma. Instead of, the corresponding forms of dark spots are completely defined by the properties of black hole gravitational field and black hole parameters like black hole mass $M$ and spin $a$.

In the nearest future it would be possible to verify modified gravity theories by observations of astrophysical black hole with international Millimetron Space Observatory \cite{Kardashev,Rudnitskiy, Likhachev, Ivanov}.

\appendix
\section{Kerr--Newman metric} \label{KN} 

The line element of the classical Kerr--Newman metric~\cite{Kerr,Newman1,Newman2, Carter68, mtw,Chandra}, describing in particular the rotating ($a\neq0$) and electrically charged ($e\neq0$) black hole, is  
\begin{equation}
	ds^2=-e^{2\nu}dt^2+e^{2\psi}(d\phi-\omega dt)^2 +e^{2\mu_1}dr^2+e^{2\mu_2}d\theta^2,
	\label{metric}
\end{equation}
where
\begin{eqnarray}
	e^{2\nu}&=&\frac{\Sigma\Delta}{A}, \quad e^{2\psi}=\frac{A\sin^2\theta}{\Sigma}, 
	\quad e^{2\mu_1}=\frac{\Sigma}{\Delta}, \quad e^{2\mu_2}=\Sigma, \quad
	\omega=\frac{2Mar}{A}, \label{omega} \\
	\Delta &= & r^2-2Mr+a^2+e^2, \quad \Sigma=r^2+a^2\cos^2\theta, \quad A=(r^2+a^2)^2-a^2\Delta\sin^2\theta, \label{A}
\end{eqnarray}
where $M$ is black hole mass, $a=J/M$ is black hole specific angular momentum (spin), $e$ is black hole electric charge, $\omega$ is frame-dragging angular velocity. 

Roots of equation $\Delta=0$ define the black hole event horizon radius $r_+$ and the Cauchy radius $r_-$:
\begin{equation}
	r_{\pm}=1\pm\sqrt{1-a^2-q^2}.
	\label{r+-}
\end{equation}
The event horizon of the Kerr-Newman black hole rotates as a solid body with (i.\,e., independent of the polar angle $\theta$) with angular velocity
\begin{equation}
	\omega_+=\frac{2Mar_+}{(r_+^2+a^2)^2}
	\label{omega+} 
\end{equation}
According to Brandon Carter equations of motion \cite{Carter68} there are the following integrals of motion: $\mu$ is  particle mass, $E$ is particle total energy, $L$ is particle azimuth angular momentum and $Q$ is the specific Carter constant, defining the non-equatorial motion. The corresponding  radial potential $R(r)$ is
\begin{equation}
	R(r) = P^2-\Delta[\mu^2r^2+(L-aE)^2+Q],
	\label{Rr} 
\end{equation}
where $P=E(r^2+a^2)-a L$. The polar potential $\Theta(\theta)$ is
\begin{equation}
	\Theta(\theta) = Q-\cos^2\theta[a^2(\mu^2-E^2)+L^2\sin^{-2}\theta].
	\label{Vtheta} 
\end{equation}

Particle trajectories depend on three parameters 
\begin{equation}
	\gamma=\frac{E}{\mu}, \quad \lambda=\frac{L}{\mu}, \quad q=\frac{\sqrt{Q}}{E}.
	\label{three} 
\end{equation}
For massless particles like photons there are two parameters: $\lambda$ and $q$. The corresponding horizontal and vertical impact parameters, $\alpha$ and $\beta$, which are viewed on the celestial sphere by a distant observer, placed at the polar angle $\theta_0$ are \cite{Bardeen73,CunnBardeen72,CunnBardeen73}): 
\begin{equation}
	\alpha =-\frac{\lambda}{\sin\theta_0}, \quad
	\beta = \pm\sqrt{\Theta(\theta_0)}.
	\label{alpha} 
\end{equation}

From astrophysical point of view (see,  e.\, g., \cite{Reynolds,Nokhrina19, Ayzenberg23})  the most probable are the cases of fast-rotating supermassive black holes with spin values close to the maximum value, $a_{\rm max}=1$.

\section{Equations of motion for test particles}  \label{Eqs}

The first order differential equations of motion in the Kerr-Newman metric, derived by Brandon Carter \cite{Carter68},  are 
\begin{eqnarray} 
	\Sigma\frac{dr}{d\tau} &=& \pm \sqrt{R(r)}, 
	\label{rmot} \\
	\Sigma\frac{d\theta}{d\tau} &=& \pm\sqrt{\Theta(\theta)}, \label{thetamot} \\
	\Sigma\frac{d\phi}{d\tau} &=& L\sin^{-2}\theta+a(\Delta^{-1}P-E), 
	\label{varphiamot}	\\
	\Sigma\frac{dt}{d\tau} &=& a(L-aE\sin^{2}\theta)+(r^2+a^2)\Delta^{-1}P,
	\label{tmot}	
\end{eqnarray}
where $\tau$ --- a proper time of the massive ($\mu\neq0$) particle or an affine parameter of massless ($\mu=0$) particle like photons.

\section{Integral equations for test particle motion} \label{Int} 

For numerical calculations are very useful the integral equations of motion (\ref{rmot})--(\ref{tmot}):
\begin{equation}\label{eq2425a}
	\quad \:\:\:	\fint\frac{dr}{\sqrt{R(r)}}
	=\fint\frac{d\theta}{\sqrt{\Theta(\theta)}}, 
\end{equation}
\begin{equation}\label{eq2425b}
	\tau=\fint\frac{r^2}{\sqrt{R(r)}}\,dr
	+\fint\frac{a^2\cos^2\theta}{\sqrt{\Theta(\theta)}}\,d\theta,
\end{equation}
\begin{equation}\label{eq25ttc}
	\phi=\fint\frac{aP}{\Delta\sqrt{R(r)}}\,dr
	+\fint\frac{L-aE\sin^2\theta}{\sin^2\theta\sqrt{\Theta(\theta)}}\,d\theta, 
\end{equation}
\begin{equation}\label{eq25ttd}
	t=\fint\frac{(r^2+a^2)P}{\Delta\sqrt{R(r)}}\,dr
	+\fint\frac{(L-aE\sin^2\theta)a}{\sqrt{\Theta(\theta)}}\,d\theta. 
\end{equation}
The integrals in equations (\ref{eq2425a})--(\ref{eq25ttd}) monotonically grow along the particle trajectory and change sign at both the radial and polar turning points:
\begin{equation}\label{eq24a}
	\int^{r_s}_{r_0}\frac{dr}{\sqrt{R(r)}}=  \int_{\theta_0}^{\theta_s}\frac{d\theta}{\sqrt{\Theta(\theta)}},
\end{equation}
where $r_s$ and $\theta_s$ are, respectively the initial and polar particle coordinates, and $r_0\gg r_{\rm h}$ and $\theta_0$ are the corresponding final particle coordinates. The more complicated case is with the one turning point in the latitude (or polar) direction, $\theta_{\rm min}(\lambda,q)$, which is a solution of the equation  $\Theta(\theta)=0$. The corresponding ordinary integrals in equation (\ref{eq2425a}) are written as 
\begin{equation}\label{eq24b}
	\int_{r_s}^{r_0}\frac{dr}{\sqrt{R(r)}}
	=\int_{\theta_{\rm min}}^{\theta_s}\frac{d\theta}{\sqrt{\Theta(\theta)}}
	+\int_{\theta_{\rm min}}^{\theta_0}\frac{d\theta}{\sqrt{\Theta(\theta)}}.
\end{equation}
We use these equations in our numerical calculations of photon trajectories, starting in the vicinity of Kerr-Newman black hole and finishing at the position of a distant observer very far from black hole.

\acknowledgments 
Author is grateful to E. O. Babichev, V. A. Berezin, Yu. N. Eroshenko, N. O. Nazarova and A. L. Smirnov for stimulating discussions.

\end{document}